\documentstyle[prl,aps]{revtex} 
\input epsf 
\begin{document} 
\draft
\title{An explanation of the ``negative neutrino mass squared'' anomaly in tritium $%
\beta$-decay based on a theory of mass } 
\author{R. L. Ingraham $\footnote{Corresponding author: Tel. (505) 646-3831,
Fax (505)646-1934; E-mail: ringraha@nmsu.edu (Richard L. Ingraham)}$} 
\address{Department of Physics, MSC 3D, New Mexico State University, P.O. Box 30001\\ 
Las Cruces, NM 88003} 
\author{G. Luna-Acosta} 
\address{Instituto de Fisica, Universidad Aut{\'{o}}noma de Puebla\\ 
Apdo. Postal J-48\\Puebla, Pue., M{\`{e}}xico 72570}

\author{J. M. Wilkes} 
\address{Air Force Research Laboratory, AFRL/DEBS\\ 
3550 Aberdeen Ave SE\\ 
Kirtland Air Force Base, NM 87117-5776, USA} 
\maketitle
\begin{abstract}
A proposed solution of the anomalous behavior of the electron spectrum near 
the endpoint of tritium $\beta$-decay is offered. It is based on a new 
theory of mass in which mass becomes a dynamical variable, and the electron
in the tritium $\beta$-decay has a narrow mass distribution. The
 predicted Kurie plots explain the main feature (``$m_{\nu}^2 
< 0\,$'') of this anomalous behavior. 
\end{abstract} 
\bigskip
\bigskip
\bigskip

Tritium beta decay shows anomalous behavior, observed in many experiments 
since 1991 \cite{Belesev,Stoeffl,Weinheimer,Holzschuh,Kawakami,Robertson,Lobashev,Wein99,Loba99}.
 Though various theoretical explanations have been offered 
\cite{Hughes,Mohapatra,Stephenson}, there is as yet no consensus
on the solution of the puzzle. The anomaly consists of two 
effects. (a) The Kurie plot lies above a certain straight line (the 
theoretical result for a massless neutrino) near the endpoint, and 
overshoots that endpoint\cite{fn1}. This is the effect that is fit by 
the infamous ``negative neutrino mass squared'' parameterization. (b) There 
is a narrow and low ``bump'' in the Kurie plot starting a few (5-10) eV 
below the endpoint\cite{fn2}.

We show that if the electrons emitted in tritium $\beta$-decay have a narrow 
mass distribution instead of the sharp mass $m_0$ assumed in the present day 
standard theory, the endpoint-overshoot effect (a) above, can be explained. 
This idea is based on a theory of mass developed by one of us; a few details
 will be given below. The electron antineutrino may be massless or have real
 mass $m_{\nu} > 0$,  though the latter case gives a better fit, see later. 
The half-width $\Delta m_e$ of the electron mass distribution
 must of course be much smaller than the central value $m_0$ ($\sim 1/2\,$ MeV, 
units: $\hbar = c = 1$), and the fits below suggest $\Delta m_e \approx m_{\nu}\,$. 

It is easy to see qualitatively why such a mass distribution would imply the 
endpoint-overshoot effect. At a value of electron momentum $p$ far from its 
endpoint value the distribution would produce a narrow scatter of events 
with electron energies both above and below $E_{0}(p)\equiv \sqrt{%
\,p^{2}\,+\,m_{0}^{2}\,}\,$. But near enough to the {\it endpoint} $%
p=p_{max}\,$ (Definition: $p_{max}\,$ is calculated from energy conservation 
for antineutrino momentum $q=0$ and mass $m_{\nu }\geq 0\,$ on the standard 
theory, electron mass sharp $=m_{0}\,$) this scatter would become 
unsymmetrical because values of electron mass $m>m_{0}\,$ would be ruled out 
by energy conservation. At $p_{max}\,$ the electron count goes to zero on 
the standard theory. But there would still be counts on the mass 
distribution theory due to values $m<m_{0}\,$. Correspondingly, the new 
Kurie plot $K(p)\,$ would coincide with the old one $K_{0}(p)\,$ for $p$ far 
from the endpoint, but at $p=p_{max}\,$, where $K_{0}(p_{max})=0\,$, $%
K(p_{max})\,$ would be positive. Thus $K(p)\,$ would overshoot the endpoint 
by an amount of order $\Delta m$. Figure 1 illustrates this 
quantitatively and in detail. 

To avoid some obvious misunderstandings we add a few words here on the theory\cite{fn3,newrli} 
underlying this proposed solution. Some further details will be given in the concluding remarks.

The unsharp mass of the electron (the mass distribution) is {\it not} due to 
decay channels---the electron is stable. Rather, mass $m$ of any elementary 
particle becomes a new dynamical variable alongside linear momentum ${\bf p}$ and energy $E$. Mass $m$ is conjugate to a new length variable $\lambda $ just as ${\bf p}$ and $E$ are to position ${\bf r}$ and time $t$, and 
roughly obeys the uncertainty relation $\Delta m$ $\Delta \lambda $ $\geq $ $1/2$ in any state. The new dimension $\lambda$ enters this theory as a microscopic length which modifies the usual point-particle causality of the 4-$D$ theory at small distances\cite{fn3}. It prevents the instantaneous action of source point on field point, and can be thought of as giving (massive) elementary particles a finite size $\sim \lambda$ \cite{fn4}. The uncertainties $\Delta m$ and $\Delta \lambda$ depend on the electron's state, here a free electron state. Thus $\Delta m$ here is expected to be different from the $\Delta m$ for, say, an electron bound in an atom or molecule. 

This theory is invariant under the Poincar{\'{e}} group in particular, so that momentum and energy are conserved. Further, it implies the usual relation $E = \sqrt{p^2 + m^2\,}\,$ for free particles.

Traditionally in physics every particle had a definite (sharp) mass in the absence of interactions leading to decays. But both theory and experiment in the last fifteen years suggest that we may be in a transitional stage in our understanding of elementary particle mass. The concept of ``particle mixing'' has entered particle physics. What this implies is that there exist particles, some observable, which have no definite mass. (We mean here: apart from the mass widths implied by possible decay channels.) Examples are the three kinds of neutrinos, $K^o$ and ${\bar {K}}^o$, the ``primed'' quarks $d'$, $s'$, and $b'$, which directly enter the weak interaction Lagrangian, etc. These particles ``oscillate'': their states are superpositions of a few other particle states with sharp mass, the ``mass (eigen)states''. In other words, the oscillating particles have (simple) mass distributions.

However, there are inconsistencies and lacunae in this transitional theory of mass. For example: (i) Among elementary particles observable as free particles there are the heavy boson $Z^o$ and the photon $\gamma$, which are mass eigenstates, and then there are the neutrinos $\nu_e$, $\nu_{\mu}$ and $\nu_{\tau}$, which are mixtures. (ii) Further, though the term ``mass (eigen)state'' has entered the literature\cite{fn5}, the corresponding ``mass operator'' for the general particle state has never been defined. (iii) There is no theory for the mixing angles, nor for the discrete and finite mass spectra we see in nature. (iv) Why are massless neutrinos necessarily only left-handed?

Suppose then that the electron has a narrow mass distribution $P(m)$. Whereas the usual free (Interaction Picture) quantum electron field appearing in the $S$-operator is\cite{Jauch}
\begin{eqnarray}
{\widehat{\psi}}_0(x) \,=\, (2\pi)^{-3/2}\,\int\,d^3p\,(m_0/E_o)^{1/2}\,e^{ip\cdot x}\,u(\bbox{p},m_0)\,a(\bbox{p}) \,+\, {\mbox{antiparticle part}}, \nonumber
\end{eqnarray}
the corresponding 4-$D$ field with the mass distribution {\it amplitude} $A(m)$ is
\begin{eqnarray}
{\widehat{\psi}}(x) \,=\, (2\pi)^{-3/2}\,\int\int\,dm\,d^3p\,(m/E)^{1/2}\,A(m)\,e^{ip\cdot x}\,u(\bbox{p},m)\,a(\bbox{p},m) \,+\, {\mbox{antiparticle part}}, \label{massq}
\end{eqnarray}
where
\begin{eqnarray}
E(\bbox{p},m) \,\equiv\, \sqrt{\,\bbox{p}^2 \,+\, m^2\,}, \quad (\,i\gamma\cdot p \,+\, m\,)u(\bbox{p},m) \,=\, 0, \quad  [\,a(\bbox{p},m),a^\dagger(\bbox{p},m)\,]_+ \,=\, \delta(\bbox{p}-\bbox{p'})\delta(m-m'), \nonumber
\end{eqnarray}
and where $P(m) \equiv |A(m)|^2\,$ and $\int_0^{\infty}\,dm\,|A(m)|^2 = 1$. Note that both ${\widehat{\psi}}_0(x)$ and ${\widehat{\psi}}(x)$ have dimension $L^{-3/2}$ ($L \equiv$ length) as they must to give a correctly normalized interaction Lagrangian.

Assume now that the electron field, Eq. (\ref{massq}), appears in the $S$-operator. Then on taking the $S$-matrix element corresponding to the decay $^3H \rightarrow \,^3He^+ + e + {\bar {\nu}}$, the electron differential spectrum ${\cal {N}}(p,m)$ (where, by definition, ${\cal {N}}(p,m) \equiv$ number of electrons per unit time with momentum $p$ and mass $m$ in the intervals $(p,p+dp)$ and $(m,m+dm)$, respectively) becomes\cite{fn6}
\begin{eqnarray}
{\cal{N}}(p,m) \,\equiv\, N(p,m)P(m), \quad N(p,m) \,\propto\, p^2qE_{\nu}. \label{Nrelations}
\end{eqnarray}
Here, $p \equiv |\bbox{p}|$ is electron momentum, $q = |\bbox{q}|$ is antineutrino momentum, $E_{\nu} \equiv \sqrt{q^2 + m_{\nu}^2\,}$ is antineutrino energy with a (real) antineutrino mass $m_{\nu} \geq 0$, and $m$ is electron mass.

Energy conservation reads
\begin{eqnarray}
Q \,\equiv\, M(\,^3H) \,-\, M(\,^3He^+) \,=\, E_e \,+\, E_{\nu}, \label{econs}
\end{eqnarray}
where (as usual) we neglect the $^3He^+$ recoil energy and for simplicity treat the case of a single $^3He^+$ final state. In the NR (nonrelativistic) case valid for tritium $\beta$-decay we can expand the electron energy as
\begin{eqnarray}
E_e \,=\, [\,p^2 \,+\, (m_0 + x)^2\,]^{1/2} \,\approx\, m_0 \,+\, T \,+\, x, \quad x \,\equiv\, m \,-\, m_0, \quad T \,\equiv\, p^2/2m_0, \label{Eapprox}
\end{eqnarray}
correct to terms linear in $x$ (assume $P(m) \approx 0$ unless $|x| << m_0$). If Eq. (\ref{Eapprox}) is used in Eq. (\ref{econs}), NR energy conservation reads
\begin{eqnarray}
T_0 \,=\, T \,+\, x \,+\, E_{\nu}, \qquad T_0 \,\equiv\, Q \,-\, m_0. \label{T0def}
\end{eqnarray}

\noindent We remark here that $T_0$ is usually called $E_0$ ($\approx$ 18,570 eV) in the experimental literature. Note that $T$ is not the kinetic energy of the electron of momentum $p$ and mass $m$ because $m_0$, not $m$, appears in the denominator. But it is convenient to use this $T$ because then the full $m$-dependence of $E_e$ to $O(x)$ appears in the single term $+ x$ in the NR case. Further, $T$ {\it is} the kinetic energy in the standard theory, and is what is used in the experimental papers (where it is usually called $E$).

After putting $q$ and $E_{\nu}$ in terms of $p$ and $x$, we get the electron differential spectrum in the form
\begin{eqnarray}
{\cal{N}}(p,m) \,\propto\, 2m_0\,T\,(\,T_0 \,-\, T \,-\,x\,)\,[\,(\,T_0 \,-\, T \,-\,x\,)^2 \,-\, m_{\nu}^2\,]^{1/2}\,P(m). \nonumber
\end{eqnarray}
It is more convenient to use a spectrum ${\cal{N}}'(T,m)$ for the number of electrons with kinetic energy $T$ between $T$ and $T + dT$, etc., so we multiply the above spectrum by $dp/dT = m_0/\sqrt{2m_0T\,}$. Changing to the convenient energy variable $y = T_0 - T\,$, we arrive at
\begin{eqnarray}
{\cal{N}}'(y,x) \,\propto\, m_0 \sqrt{2m_0\,(T_0 - y)\,}\,(\,y \,-\,x\,)\,[\,(\,y \,-\,x\,)^2 \,-\, m_{\nu}^2\,]^{1/2}\,P(x)\,dy dx, \label{Nprimeyx}
\end{eqnarray}
where we wrote ${\cal{N}}'(y,x)\equiv {\cal{N}}'(T,m)$ and $P(x) \equiv P(m)$ for notational simplicity. Note that since momentum $p$ (or equivalently, kinetic energy $T = p^2/2m_0\,$) is measured in the experiments {\it but not mass m}, the differential spectrum of interest is
\begin{eqnarray}
{\cal{N}}'(y) \,\equiv\, \int\,{\cal{N}}'(y,x)dx. \label{Nprimey}
\end{eqnarray}

{\it Integrated spectrum}. In the experiments{\cite{Belesev,Stoeffl,Weinheimer,Holzschuh,Kawakami,Robertson,Lobashev,Wein99,Loba99}}, when a Kurie plot is given, it is the integrated spectrum (differential spectrum integrated in $T$ from some $T$ to the endpoint). Hence we consider the integral of ${\cal{N}}'(y)$ from the endpoint $y = 0$ to some $y > 0$. The only values of $y$ of interest are in the endpoint region $y << T_0$, so we can put $T_0 - y \rightarrow T_0$ in Eq. (\ref{Nprimeyx}). Then from Eqs. (\ref{Nprimeyx}) and (\ref{Nprimey}):
\begin{eqnarray}
{\cal{N}}^{int}(y) \,\propto\, m_0\,\sqrt{2m_0\,T_0\,}\,\int_{-n\Delta m}^{y-m_{\nu}}dx\,P(x)\,\int_{x+m_{\nu}}^y\,dy'(\,y' \,-\,x\,)\,[\,(\,y' \,-\,x\,)^2 \,-\, m_{\nu}^2\,]^{1/2}. \label{Nint}
\end{eqnarray}
The limits were determined from energy conservation, Eq. (\ref{T0def}), on the assumption that the $y'$-integral is done first for a given value of $x$. In the $x$-integral, $\Delta m$ is the half-width of the mass distribution, and $n$ is some small integer (2 or 3) such that $P(x)$ is considered effectively zero for $x \leq -n\Delta m$.

Remarkably, the $y'$-integral in Eq. (\ref{Nint}) can be done analytically. Then we define a normalized\cite{fn7} integrated Kurie plot $K(y)$ by dividing out some factors and taking the cube root of the resulting integral $I(y)\,$:
\begin{mathletters}
\label{IyandK}
\begin{eqnarray}
I(y) \,\equiv\, \int_{-n\Delta m}^{y-m_{\nu}}dx\,P(x)\,&&[\,(\,y \,-\,x\,)^2 \,-\, m_{\nu}^2\,]^{3/2},\quad m_{\nu} - n\Delta m < y << T_0,\label{IyandKa}
\end{eqnarray}
\begin{eqnarray}
K(y) \,\equiv\, [I(y)]^{1/3}. \label{IyandKb}
\end{eqnarray}
\end{mathletters}
The integral $I(y)\,$ must be done numerically, and we chose $P(x)$ to be the gaussian
\begin{eqnarray}   
P(x) \,=\,  \frac{1}{\sqrt{2\pi\,}\,\sigma}\,{\mbox{exp}}\Big(-\frac{x^2}{2\sigma^2} \Big)\label{Px}
\end{eqnarray}  
for the numerical integration, in which case we set $\Delta m = \sigma$, the standard deviation of the gaussian. But see the concluding remarks.

There are two mass parameters available, $\sigma$ and $m_{\nu}$. It is convenient to non-dimensionalize Eqs. (\ref{IyandK}) and (\ref{Px}) by using one of them. We chose $m_{\nu}$, and denoted the dimensionless quantities by carets:
\begin{eqnarray}
{\widehat{x}} \equiv x/m_{\nu}, \quad {\widehat{y}} \equiv y/m_{\nu}, \quad {\widehat{\sigma}} \equiv \sigma/m_{\nu}, \quad {\widehat{T}}_0 = T_0/m_{\nu}; \qquad
{\widehat{P}}({\widehat{x}}) \equiv m_{\nu}\,P(x), \quad {\widehat{I}}({\widehat{y}}) \equiv I(y)/m_{\nu}^3, \quad {\widehat{K}}({\widehat{y}}) \equiv K(y)/m_{\nu}. \nonumber
\end{eqnarray}
Then the quantity plotted in Fig. 1  is ${\widehat{K}}({\widehat{y}}) \equiv [{\widehat{I}}({\widehat{y}})]^{1/3}$, where
\begin{eqnarray}
{\widehat{I}}({\widehat{y}}) \,\equiv\, \int_{-n{\widehat{\sigma}}}^{{\widehat{y}}-1}d{\widehat{x}}\,{\widehat{P}}({\widehat{x}})\,[\,(\,{\widehat{y}} \,-\,{\widehat{x}}\,)^2 \,-\, 1\,]^{3/2}, \qquad 1 - n{\widehat{\sigma}} < {\widehat{y}} << {\widehat{T}}_0, \label{finalint}
\end{eqnarray}
for various values of the ratio ${\widehat{\sigma}} \equiv \sigma/m_{\nu}$.

Note that in the sharp mass limit ${\widehat {P}}({\widehat {x}})$ goes to a delta function, ${\widehat{I}}({\widehat{y}})$ becomes $({\widehat{y}}^2 - 1)^{3/2}$, and
\begin{eqnarray}
{\widehat{K}}({\widehat{y}}) \rightarrow (\,{\widehat{y}}^2 \,-\, 1\,)^{1/2} \,\approx\, {\widehat{y}} \,-\, 1/2{\widehat{y}}, \quad{\mbox{for}} \quad {\widehat{y}} >> 1. \nonumber
\end{eqnarray}
This is a straight line of slope $+1$ (in ${\widehat{y}}\,$) far from the endpoint, but near the endpoint it curves down and vanishes at ${\widehat{y}} = 1$ below the endpoint, the well-known result for a massive neutrino in the standard theory.
 (For $m_{\nu} = 0\,$, the corresponding plot of $K(y)$ would be a straight line all the way to $y=0$.)

In the present, mass unsharp theory for the case $m_{\nu} = 0\,$ we must return to $K(y)$,
 Eq. (\ref{IyandKb}), with $m_{\nu}\,$ set to zero. This can be non-dimensionalized by dividing
 all quantities by suitable powers of $\sigma$. We have omitted this graph here; the Kurie plot
 lies approximately half way between the plotted curves for ${\widehat{\sigma}} = 1.0$ and $1.5$
 shown in the figure.
\begin{figure}
\epsfysize=3.4truein
{\centerline {\epsffile{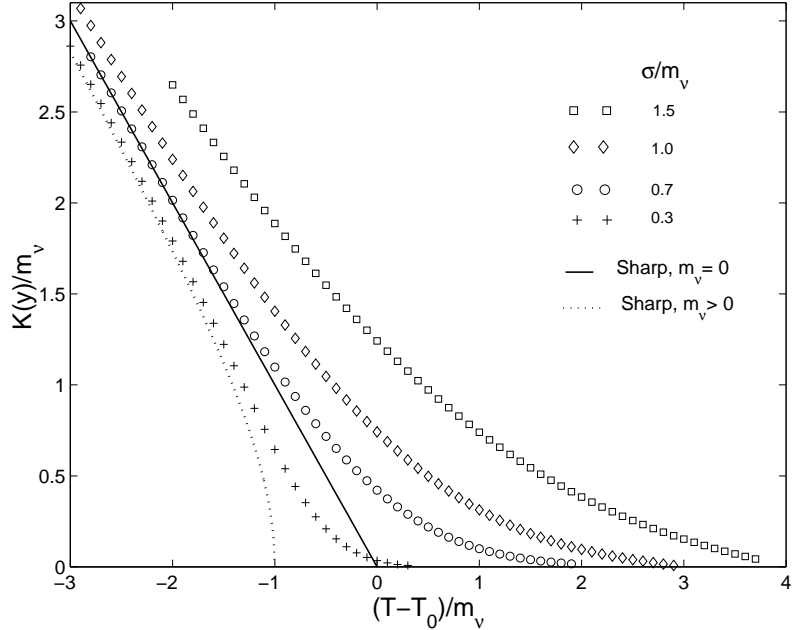}}}
\caption{The non-dimensionalized
Kurie plots ${\widehat{K}}({\widehat{y}}) \equiv K(y)/m_{\nu}$ against
$-{\widehat{y}} \equiv (T - T_0)/m_{\nu}$ (the latter for ease of comparison with
experimental results), for $\widehat{\sigma}=0.3,0.7,1.0$,and $1.5$ from left to right.
Also shown are the Kurie plots of ${\widehat{K}}_0({\widehat{y}})$ in the sharp mass 
case for massive and massless neutrino.}
\label{fig1}
\end{figure}

{\it Discussion of the results}. Fig. 1 shows the non-dimensionalized Kurie plot for the cases 
${\widehat{\sigma}} = 0.3, 0.7, 1.0$, and $1.5$. For ease of comparison with experimental results,
 however, we have plotted $-{\widehat{y}} \equiv (T - T_0)/m_{\nu}$ on the horizontal axis. Also
 shown are the Kurie plots in the sharp mass case: ${\widehat{K}}_0({\widehat{y}}) = 
(\,{\widehat{y}}^2 \,-\, 1\,)^{1/2}$ for $m_{\nu} > 0$ and the straight line 
${\widehat{K}}_0({\widehat{y}}) = -{\widehat{y}}$ for $m_{\nu} = 0\,$ (the latter is the dimensional
 Kurie plot $K_0(y) = -y$ arbitrarily normalized with an $m_{\nu} \not = 0\,$). $K(y)$ stays
 above the straight line and overshoots the endpoint $y = 0$ ($\,T = T_0\,$) by about $2m_{\nu},3m_{\nu}$,
 or $4 m_{\nu}\,$ in the cases ${\widehat{\sigma}} = 0.7$, $1.0$, and $1.5$, respectively. If one knew or
 guessed a value for $m_{\nu}$, this would give the predicted overshoot quantitatively. The comparison 
with experiment would be clearer if the data at and just beyond the endpoint were cleaner, as previously
 mentioned\cite{fn1}. The curves for the 
 values ${\widehat{\sigma}} < 0.3$ lie below the straight line and do not overshoot 
 the point $T = T_0$.    There is no indication in any of these curves of effect (b), the ``bump".  
The plots in the Figure resemble the experimental plots in Refs. [1,2,3,5,8, 
 and 9] in a general way.  In particular, the plot for ${\widehat{\sigma}} = 0.7$
 appears to be an especially good fit to the experimental points in Fig. 1 
 of the recent work [9].\\

{\it Concluding remarks}. 
 (a) The gaussian, Eq.(10) was chosen arbitrarily as a typical
 narrow mass distribution, but calculations using other mass distribution functions indicate that
 the Kurie plots do not depend sensitively on the shape of the distribution so long as it is 
narrow.  We should mention that the underlying theory\cite{fn3} is capable of predicting this
 mass distribution uniquely. Further, recent work\cite{newrli} suggests how the effective 4-$D$ field
 ${\widehat{\psi}}(x)$, Eq. (\ref{massq}), arises from the natural 5-$D$ free quantum field 
${\widehat{\Psi}}(x,\lambda)$ of this 5-$D$ theory. However, these ideas are not yet in final
 form and anyway would be too lengthy to give here. (b) The possible objection that the uncertainty
 in the electron mass found in particle property tables is too small to explain this 
$^3H$ $\beta$-decay anomaly is no valid criticism of this proposal. To repeat: if mass becomes a
 dynamical variable, its uncertainty depends on the state, and may vary wildly with that state. 
The value $0.3$ ppm for the uncertainty quoted in the Tables\cite{fn8} comes from the most precise
 atomic measurements. There is no reason for the mass width to be the same as that for a free 
electron especially if this ``free'' electron is acted on by a self-force. (A self-force was found 
essential in the classical theory to give a nonsingular self-interaction with all the correct 
properties\cite{fn3} and in the quantum theory to enable the prediction of mass spectra for
elementary particles\cite{newrli}). To our knowledge no experiment measuring the mass
 width of a free electron, such as emerges in tritium $\beta$-decay, has ever been performed, 
though it would be simple in principle.(c) We can give an upper bound on $\sigma =\Delta m_e$ on 
the basis of our results as follows. We predict $\sigma =  \widehat{\sigma}m_{\nu} \approx 0.7 m_{\nu}$ (see above).  If $m_{\nu} < 1$ eV for example, then
 $\sigma < 0.7$ eV, or $ < 1.4$ ppm.
\\
\acknowledgements{G.A. L-A. acknowledges support from CONACYT, Grant No. 26163-E.}


\begin{references}
\bibitem{fn1}  Comparison with experiment is complicated by large error bars 
and subtraction of background counts around the endpoint. The latter is 
especially worrisome here because real electron counts beyond the endpoint 
predicted by this theory (see just below) may have been automatically 
considered background and deleted.
\bibitem{fn7}  So normalized, our $K(y)$ has dimensions of $M$ or $L^{-1}$ ($\hbar = c =1$), so cannot be directly compared in size to experimental Kurie plots, which usually display $(counts/s)^{1/3}$.
\end{references}
\end{document}